\documentclass[letterpaper, 10 pt, conference]{ieeeconf}
\newtheorem{lemma}{Lemma}{}
\newtheorem{remark}{Remark}{}
\usepackage{amsmath}
\usepackage{graphicx, amssymb}
\usepackage[dvips]{epsfig}
\usepackage{color}

\IEEEoverridecommandlockouts
\newcommand{\carrew} {\hfill $\Box$}
\newcommand{\carre}{\begin{flushright} \rule{2mm}{2mm} \end{flushright}}
\newcommand{\jgrv}[1]{{\color{black} #1}}

\title{Solution of matching equations of IDA-PBC by Pfaffian differential equations
}

\author{M. Reza J. Harandi$^{1}$ and  Hamid. D. Taghirad$^{1}$
	\thanks{$^{1}$M. Reza J. Harandi is PhD student in the Department of Electrical Engineering,
		K. N. Toosi University of Technology, Tehran,  Iran
		{\tt\small jafari@email.kntu.ac.ir,  taghirad@kntu.ac.ir}}%
}
\begin{document}
	\maketitle
	\thispagestyle{empty}
	\pagestyle{empty}

\begin{abstract}
Finding the general solution of partial differential equations
(PDEs) is essential for controller design in newly developed
methods. Interconnection and damping assignment passivity based
control (IDA-PBC) is one of such methods in which the solution to
corresponding PDEs \jgrv{which are called matching equations,} is
needed to apply it in practice. In this paper, \jgrv{these matching
	equations} are transformed to corresponding Pfaffian differential
equations. Furthermore, it is shown that upon satisfaction of the
integrability condition, the solution to the corresponding third-order Pfaffian differential equation may be obtained quite easily.
The method is applied to the PDEs of IDA-PBC  in some benchmark
systems such as Magnetic levitation system, Pendubot, and
underactuated cable driven robot to verify its applicability.
\end{abstract}
\section{Introduction}
\label{intro}

Solving partial differential equations (PDEs) is one of the most
challenging problems in mathematics. This issue is more crucial when
the general solution is required while no boundary condition exists.
One of the applications of such problems in control engineering is
where the controller design in some methods is based on the solution
of some PDEs. Interconnection and damping assignment passivity based
control (IDA-PBC) is one of the well-known methods whose application
is restricted to the prohibitive task of finding general solution of
PDEs~\cite{ortega2002interconnection}.

Port-Controlled Hamiltonian (PCH) is a the general methods of
modeling physical systems by determination of a Hamiltonian function
together with interconnection and damping
matrices~\cite{qureshi2015}. One method to stabilize PCH systems is
classical passivity based control where at priory the Hamiltonian as
the storage function shall be assigned to the system and then a
suitable controller shall be designed to minimize the storage
function~\cite{ortega2004interconnection, donaire2017robust}. In order to rectify some
of the technical issues of this method, a second class of solution
was proposed, in which, instead of fixing the closed-loop storage
function, the desired structure of the closed-loop system is
assigned~\cite{muhammad2012passivity}. Interconnection and damping
assignment~\cite{ortega2002interconnection} and controlled
Lagrangian~\cite{bloch2000controlled} are examples of such
rectification. The energy function to be assigned is found by the
solutions of a set of PDEs that is called matching equations~\cite{gupta2019controller}. This
energy function is then used to design the stabilizing controller
for the system~\cite{franco2019ida}. Since these PDEs do not
have boundary conditions and the solution shall acquire their
minimum value at the desired equilibrium point, obtaining the
general solution of PDEs is required, which is usually a prohibitive
task.

This problem is the focus of attention of many researches.
In~\cite{acosta2005interconnection}, a method for mechanical systems
with one degree of underactuation has been developed. In this work,
it is shown that upon satisfying some conditions, potential and
kinetic energy PDEs may be solved easily.
Reference~\cite{viola2007total}, has striven to simplify the kinetic
energy PDEs of underactuated mechanical systems by coordinate
transformation, while in~\cite{acosta2009pdes}, the matching
equations are replaced by algebraic inequalities. Constructive
IDA-PBC for PCH systems has been introduced
in~\cite{nunna2015constructive} by which the PDEs are replaced by
algebraic equations. In~\cite{donaire2016simultaneous} simultaneous
IDA-PBC was proposed in which using dissipative forces a more
general version of  kinetic energy PDEs of mechanical systems was
derived. References \cite{borja2015shaping, donaire2015shaping,
mehra2017control}, are some representative works
that have focused on this issue. Generally, these works may be
separated into two categories, some of them include a very special
class of PCH systems while the corresponding matching equations can
be solved quite easily. On the contrary, other methods are
applicable to a large class of systems while performing their
solution in most cases is as hard as solving the original PDEs.

In this paper, we utilize one of the less focused methods proposed
in the literature~\cite{sneddon2006elements}, to derive the general
solution of a PDE. In this reference, it is shown that a first-order
PDE with $n$ variables is equivalent to $n$ Pfaffian differential
equations. By this means, finding suitable solution of the PDE is
simplified to find the solution to its corresponding Pfaffian
differential equations. Generally, solving this form of differential
equations is not an easy task. However, for a third-order Pfaffian
equation that satisfies a certain condition, several methods may be
employed to derive the solution. Therefore, for a PDE with three
variables, one may derive a Pfaffian differential equation and try
to transform the equations such that the required condition is
satisfied. By this means, the solution could be derived easily. Note
that one of the most important differences of this method to other
proposed methods like characteristic methods detailed in~\cite[Ch.
3]{evans2010partial}, is that the stringent requirement to know the
boundary conditions in order to compute the solution of PDE is
released.

In what follows, details of this method is introduced, and it is
applied to solve some benchmark systems. Notice that the basic
mathematics of this work is borrowed
from~\cite{sneddon2006elements}, and in this paper we aim to show
the applicability of this method in general and to use it to solve
the challenging PDE of an underactuated cable driven robot
introduced in~\cite{harandi2019point}.

\section{Background mathematics}
One of the well-known methods for stabilization of dynamical systems
is IDA-PBC~\cite{ortega2004interconnection}.  In the following, we
briefly introduce this method and investigate the PDEs arisen in
this method for some benchmark systems.

Consider a class of port-controlled Hamiltonian systems with dynamic
formulation of the following form
\begin{equation}\label{6}
\dot{x}=\big(J(x)-R(x)\big)\nabla H+g(x)u,
\end{equation}
where $x\in\mathbb{R}^n$ denotes the states of the  system,
$u\in\mathbb{R}^m$ denote the input, $J(x)=-J^T(x)$ and
$R(x)=R^T(x)\geq0$ are the interconnection and damping matrices
respectively, and $H(x):\mathbb{R}^n\to \mathbb{R}$ denoted the
total stored energy in the system. The IDA-PBC method relies on
matching the system (\ref{6}) with a generalized Hamiltonian
structure
\begin{equation}\label{7}
\dot{x}=\big(J_d(x)-R_d(x)\big)\nabla H_d(x)
\end{equation}
in which $H_d(x)$ is continuously differentiable desired storage
function which is (locally) minimum at the desired equilibrium point
$x^*$, while $J_d(x)=-J_d(x)$ and $R_d(x)=R_d^T(x)\geq 0$ represent
desired interconnection and damping terms, respectively.

Assume that matrix $g^\perp (x):\mathbb{R}\to\mathbb{R}^{n-m}$ which
is the full rank left annihilator of $g(x)$ and $J_d,R_d$ and $H_d$
such that the following equation is satisfied:
\begin{equation}\label{3}
\resizebox{.92\hsize}{!}{$g^\perp (x)\big(J(x)-R(x)\big)\nabla H(x)=
    g^\perp (x)\big(J_d(x)-R_d(x)\big)\nabla H_d(x)$}
\end{equation}
This equation results from matching the systems (\ref{6}) and
(\ref{7}). If this condition holds, then the open-loop system
(\ref{6}) with the feedback
\begin{align}
u(x)=&\, (g^T g)^{-1}g^T\times\Big(\big(J_d(x)-R_d(x)\big)\nabla
H_d(x)\nonumber\\&-\big(J(x)-R(x)\big) \nabla H(x)\Big) \label{5}
\end{align}
may be written in form of (\ref{7}), whose $x^*$ is a (locally)
stable equilibrium point~\cite{nunna2015constructive}.

As a special case of using this method in mechanical systems,
consider the general dynamic formulation of a robot in
port-controlled Hamiltonian form as:
\begin{equation}
\label{1a}
\begin{bmatrix}
\dot{q} \\ \dot{p}
\end{bmatrix}
=\begin{bmatrix}
0_{n\times n} & I_n \\ -I_n & 0_{n\times n}
\end{bmatrix}
\begin{bmatrix}
\nabla_q H \\ \nabla_p H
\end{bmatrix}
+\begin{bmatrix}
0_{n\times m} \\ G(q)
\end{bmatrix}
\tau
\end{equation}
where $H(q,p)=\frac{1}{2}p^TM^{-1}(q)p+V(q)$ is total energy of the
system as the sum of kinetic and potential energy, $q,p\in R^n$
denote generalized position and orientation, $M^T(q)=M(q)>0$ denotes
the inertia matrix and $G(q)\in \mathbb{R}^{n\times m}$ is the input
coupling matrix. suppose that the desired storage function is set to
$H_d=\frac{1}{2}p^TM_d^{-1}(q)p+V_d(q)$ in which
\begin{equation}
\label{1.5} q^*=\text{arg min}\, V_d(q)
\end{equation}
where $q^*$ is desired equilibrium point. Desired structure of the
closed-loop system is considered as follows
\begin{equation}
\label{2a}
\begin{bmatrix}
\dot{q} \\ \dot{p}
\end{bmatrix}
=
\begin{bmatrix}
0_{n\times n} & M^{-1}M_d \\ -M_dM^{-1} & J_2-GK_vG^T
\end{bmatrix}
\begin{bmatrix}
\nabla_q H_d \\ \nabla_p H_d
\end{bmatrix}
\end{equation}
in which $J_2(q,p)\in\mathbb{R}^{n\times n}$ is an skew-symmetric
matrix.  For this representation the control law may be derived as
follows
\begin{equation}
\label{3a}
\begin{array}{c}
\tau=(G^TG)^{-1}G^T\bigg(\nabla_q V-M_dM^{-1}\nabla_q V_d+\nabla_q K\\
-M_dM^{-1}\nabla_q K_d+(J_2-GK_vG^T)\nabla_p H_d\bigg)
\end{array}
\end{equation}
in which, $M_d,V_d$ shall satisfy the following PDEs
\begin{align}
\label{4a}
\resizebox{.97\hsize}{!}{$G^\bot (q)\{\nabla_q \big(p^TM^{-1}(q)p\big)-M_dM^{-1}(q)\nabla_q \big(p^TM_d^{-1}(q)p\big)$}&\nonumber\\+2J_2M_d^{-1}p\}=0.&
\end{align}
This is called the kinetic energy PDE (KE-PDE), while the potential
energy PDE (PE-PDE) may be written as follows:
\begin{equation}
\label{5a}
G^\bot (q)\{\nabla_q V(q)-M_dM^{-1}\nabla_q V_d(q)\}=0
\end{equation}
in which, $G^\bot$ is left annihilator of $G$ (i.e. $G^\bot G=0$).

\section{Main results}\label{s3}
In this section, let us introduce the method proposed in
\cite[Ch.2.3]{sneddon2006elements} to solve first-order PDEs that
may arise in various areas of control engineering. In this method, a
PDE with $n$ independent variables is converted to $n$ Pfaffian
differential equations which are generally in the following form:
\begin{equation}
\displaystyle\sum_{i=1}^{n}f_i(x_1,...,x_n)d x_i=0.
\end{equation}
Let us restate Theorem~3 of \cite[Ch.2.3]{sneddon2006elements} for
ease of use in here. We suggest reading the proof and examples in
the main reference.\\
\textbf{Theorem}~\cite{sneddon2006elements}: If
$\phi_i(x_1,\dots,x_n,z)=c_i$ where $i=1,\dots,n$, are independent
solutions of the equations
\begin{align}
\label{1}
\frac{d x_1}{P_1}=\frac{d x_2}{P_2}=\dots=\frac{d x_n}{P_n}=\frac{d
    z}{R},
\end{align}
then for the arbitrary function $\Phi$,   $\Phi(\phi_1,\dots,\phi_n)=0$,
forms a general solution of the following partial differential
equation
\begin{equation}
\label{2} P_1\frac{\partial z}{\partial x_1}+ P_2\frac{\partial
z}{\partial x_2}+\dots+P_n\frac{\partial z}{\partial x_n}=R,
\end{equation}
in which, $P_i$s and $R$ are functions of $x_1,\dots,x_n,z$.\carrew

To find a general solution for Pfaffian equations (\ref{1}) is still
a prohibitive task, while it is much easier than that of the
corresponding PDE. In this book, some primary methods are proposed
to solve (\ref{1}). Moreover, for a Pfaffian differential equation
with $n=3$, i.e.
\begin{equation}
\label{pf}
Pdx_1+Qdx_2+Rdx_3=0,
\end{equation}
it is shown that if the following condition holds:
\begin{equation}
\label{cr}
X^Tcurl(X)=0,
\end{equation}
with $X=[P,Q,R]^T$, then the problem turns to an exact differential
equation which may be easily solved by direct integration. In this
case, several method is proposed to derive the solution of
(\ref{pf}). 
By this means, in order to solve (\ref{2}) with $n=3$, one may
derive a Pfaffian differential equation such that condition
(\ref{cr}) holds. In the following, various examples are given to
show the applicability of this method, and different solution
methods to the corresponding Pfaffian equations are examined in
details. One of the these methods which is proposed
in~\cite[Ch. 1]{sneddon2006elements} is summarized here.\\
\\
Assume that Pfaffian equation (\ref{pf}) satisfy condition
(\ref{cr}).\\
Stage 1: Assume that $x_3$ is constant. The solution of
$$Pdx_1+Qdx_2=0$$ is $U(x_1,x_2,x_3)=C$. Define
$$\mu:=\frac{1}{P} \frac{\partial U}{\partial x_1}=\frac{1}{Q} \frac{\partial U}{\partial x_2}$$
\\
Stage 2: Define
$$K:=\mu R-\frac{\partial U}{\partial x_3}$$
\\
Stage 3: Parameterize $K$ such that $K=K(U,x_3)$.
\\
Stage 4: Solve $dU+Kdx_3=0$.
\\
Stage 5: Then the solution is
$$\phi(U,x_3)=\phi(U(x_1,x_2,x_3),x_3)=C.$$

Before giving the details of the solutions, let us introduce the
following useful lemma.
\begin{lemma}\label{l1}\normalfont
    Consider PDE (\ref{2}) and assume that $P_i$s and $R$ are only
    functions of independent variables $x_i$s. Then,

    a) The functions $z-\phi_i(x_1,...,x_n)=c_i,i\in \{1,...,n-1\}$ are
    the homogeneous solutions of this PDE  if $\phi_i$s are solutions of
    the first $n-1$ Pfaffian equations.

    b) Non-homogeneous solution is derived by equalizing the last term
    to other terms in Pfaffian equations.\carrew
\end{lemma}
\textit{proof}: a) Assume that $\phi_i(q_1,...,q_n)=c_i$ are the
solutions of first $n-1$ Pfaffian equations. Since this equations
are independent of $z$, therefore, $z-\phi_i(q_1,...,q_n)=c_i$ are
also the solutions of Pfaffian equations. Notice that they are
homogeneous solutions of PDE, because they satisfy the following
equations which are related to homogeneous part of PDE
$$\frac{d x_1}{P_1(x_1,...,x_n)}=\dots=\frac{d x_n}{P_n(x_1,...,qx_n)}=\frac{d z}{0}$$

b) The proof of this part is clear. Non-homogeneous solution of PDE
(\ref{2}) corresponds to its special solution that depends on both
the left-- and right--hand sides. Hence, this is derived based on
the last term of (\ref{1}).
\carre

In the next section the IDA-PBC method is applied to some benchmark
systems. Note that more examples are given
in~\cite{harandi2020solution}.

\section{Benchmark examples}
In what follows, we apply the proposed method to the PDEs arisen in some benchmark
systems. Since the main objective is to detail different methods
to derive the solution of the arisen PDE in controller design, no
simulation study is given.
\begin{figure}[t]
    \center
    \includegraphics[scale=.6]{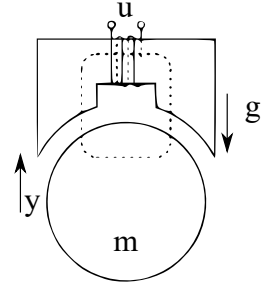}%
    \centering \caption{ Schematic of magnetic levitation system. }
    \label{sh0}
\end{figure}
\subsection{Magnetic levitation system}
This system consist of an ferric ball hovered above a magnetic field
created by an electromagnet. The schematic
of system is illustrated in Fig.~\ref{sh0}. Consider $\lambda$
as the flux generated by the magnet and $\theta$ as the distance of
the center of mass of the ball to its nominal position. It is shown
in~\cite{ortega2001putting} that the system may be represented in
PCH form as follows:
\begin{align}
\dot{x}=\begin{bmatrix}
-r & 0 & 0\\
0 & 0 & 1 \\
0 & -1 & 0
\end{bmatrix}
\frac{\partial H}{\partial x}+\begin{bmatrix}
1 \\ 0 \\ 0
\end{bmatrix} u,
\label{8}
\end{align}
in which, $x=[\lambda,\theta,m\dot{\theta}]^T$ and $r$ represent coil resistance while the Hamiltonian
function is given by
$$H(x)=\frac{1}{2k}(1-x_2)x_1^2+\frac{1}{2m}x_3^2+mgx_2.$$
Let us consider stabilizing the equilibrium point
$x^*=[\sqrt{2kmg},x_2^*,\\0]^T$ for this system. It is shown in
\cite{ortega2001putting} that without modification of
interconnection matrix, it is not possible to stabilize $x^*$.
Hence, the following interconnection matrix is considered in here:
$$\begin{bmatrix}
0 & 0 & -\alpha\\
0 & 0 & 1 \\
\alpha & -1 & 0
\end{bmatrix}.$$
Furthermore, with $R_d=R$, the matching equation (\ref{3}) yields to
\begin{align}
&K_3(x)=0\nonumber\\
&\alpha K_1(x)-K_2(x)=-\frac{\alpha}{k}(1-x_2)x_1 \label{9}
\end{align}
where it is assumed that $H_d=H+H_a$ and
\begin{equation}
\label{9.5}
[K_1,K_2,K_3]^T:=\frac{\partial
    H_a}{\partial x}=\left [\frac{\partial H_a}{\partial x_1},\frac{\partial
    H_a}{\partial x_2},\frac{\partial H_a}{\partial x_3}\right ]^T .
\end{equation}
The PDE represented by (\ref{9}) shows that $H_a$ is independent of
$x_3$. Using the proposed method, this PDE is equivalent to the
following Pfaffian equations:
\begin{align*}
\frac{dx_1}{\alpha}=\frac{dx_2}{-1}=\frac{dH_a}{-\frac{\alpha}{k}(1-x_2)x_1}
\end{align*}
If $\alpha\neq 0$ one may rewrite it as
\begin{align}
\frac{dx_1}{1}=\frac{dx_2}{-\beta}=\frac{dH_a}{-\frac{1}{k}(1-x_2)x_1}
\label{10}
\end{align}
with $\beta=1/\alpha$. Consider that $\beta=-c_1x_1-c_2x_2-c_3$ with
$c_i$s as arbitrary constants, and substitute it in (\ref{10}):
\begin{align}
\frac{dx_1}{1}=\frac{dx_2}{c_1x_1+c_2x_2+c_3}=\frac{dH_a}{-\frac{1}{k}(1-x_2)x_1}
\label{14}
\end{align}
First, derive non-homogeneous solution for the system by using
Lemma~\ref{l1}. The strategy is to derive a Pfaffian equation
satisfying (\ref{cr}). With some manipulation one may show that
equation (\ref{14}) is equal to the following:
\begin{equation*}
\frac{dH_a-\frac{x_1}{c_2k}dx_2}{-\frac{x_1}{k}-\frac{c_1x_1^2}{c_2k}-
\frac{c_3x_1}{c_2k}}=\frac{dH_a-\frac{x_1}{c_2k}dx_2-\frac{x_2}
{c_2k}dx_1}{-\frac{x_1}{k}-\frac{c_1x_1^2}{c_2k}-\frac{c_3x_1}{c_2k}-\frac{x_2}{c_2k}}
\end{equation*}
In this representation, the term $\frac{x_1x_2}{k}$ was omitted in
the denominator of $dH_a$ and then the term $-\frac{x_2}{c_2k}dx_1$
was added to satisfy condition (\ref{cr}).

In order to eliminate $\frac{x_2}{c_2k}$ from right hand side of
this equation, let us add it with $\frac{1}{c_2^2k}dx_2$. One may
verify that the denominator of this new term depends only to $x_1$.
Hence, using the first term of (\ref{14}), it is possible to omit
the remaining terms. By this means, the following Pfaffian
differential equation is derived:
\begin{align}
&\resizebox{.98\hsize}{!}{$dH_a-\frac{x_1}{c_2k}dx_2-\frac{x_2}{c_2k}dx_1+\frac{1}{c_2^2k}dx_2+
\frac{x_1}{k}dx_1+\frac{c_1x_1^2}{c_2k}dx_1$}\nonumber\\&+\frac{c_3x_1}
{c_2k}dx_1-\frac{c_1x_1}{c_2^2k}dx_1-\frac{c_3}{c_2^2k}dx_1=0.
\label{15}
\end{align}
This equation satisfies condition (\ref{cr}), and is separable
in the following form:
\begin{align*}
&(dH_a)-\Big(\frac{x_1}{c_2k}dx_2+\frac{x_2}{c_2k}dx_1\Big)+
\Big(\frac{1}{c_2^2k}dx_2\Big)+\Big(\frac{x_1}{k}dx_1\\
&+\frac{c_1x_1^2}{c_2k}dx_1+\frac{c_3x_1}{c_2k}dx_1-
\frac{c_1x_1}{c_2^2k}dx_1-\frac{c_3}{c_2^2k}dx_1\Big)=0.
\end{align*}
Therefore, one may find the following solution
$$H_a=\frac{x_1x_2}{c_2k}-\frac{x_2}{c_2^2k}-\frac{x_1^2}{2k}-\frac{c_1x_1^3}{3c_2k}-\frac{c_3x_1^2}{2c_2k}+\frac{c_1x_1^2}{2c_2^2k}-\frac{c_3x_1}{c_2^2k}.$$
Furthermore, by using Lemma~\ref{l1} the homogeneous solution is
derived from the following equation as
$$(c_1x_1+c_2x_2+c_3)dx_1-dx_2=0.$$
This equation need an integration factor $\mu$ which satisfies the
following relation
$$\frac{\partial \mu}{\partial x_1}+(c_1x_1+c_2x_2+c_3)
\frac{\partial \mu}{\partial x_2}=-c_2\mu.$$ Hence, using the
proposed Theorem, this is equivalent to
$$\frac{dx_1}{1}=\frac{dx_2}{c_1x_1+c_2x_2+c_3}=\frac{d \mu}{-c_2\mu}.$$
By considering the first and last terms, the solution may be given
as $\mu=e^{-c_2x_1}$. Hence, homogeneous solution of (\ref{14}) is
given as:
$$H_a=\phi\Big((\frac{c_1}{c_2}x_1+x_2+
\frac{c_3}{c_2}+\frac{c_1}{c_2^2})e^{-c_2x_1}\Big).$$
In which, the function $\phi$ and the constants $c_i$s shall be
determined such that $x^*$ becomes stable.
\begin{remark}\normalfont
References~\cite{ortega2001putting,nunna2015constructive}, state
    that $\theta$ shall remain in the interval of $(-1,\infty)$ while
    this limitation is released in our proposed solution. Note that
    using the method proposed in~\cite{tee2009barrier} based on control
    barrier functions, one may define $c_i$s such that this constraint
    is satisfied too. Furthermore, for the solution given
    in~\cite{ortega2001putting} it is assumed that $\alpha$ is constant.
    This limiting assumption is also released in the proposed solution
    given in this paper.
\end{remark}

\subsection{Micro electro--mechanical optical switch}
Another benchmark example is this field is the optical switching
system with the following PCH
model~\cite{borja2015shaping,borovic2004control}:
\begin{align*}
\dot{x}=\begin{bmatrix}
0 & 1 & 0 \\
-1 & -b & 0 \\
0 & 0 & -\frac{1}{r}
\end{bmatrix}\nabla H(x)+\begin{bmatrix}
0 \\ 0 \\ \frac{1}{r}
\end{bmatrix}u
\end{align*}
whose energy function is given by:
$$H(x)=\frac{1}{2m}x_2^2+\frac{1}{2}a_1x_1^2+\frac{1}{4}a_2x_1^4+
\frac{x_3^2}{2c_1(x_1+c_0)},$$ where $b,r>0$ are resistive
constants, $a_1,a_2>0$  are spring terms, $c_0,c_1>0$ are capacitive
elements and $m$ denotes the mass of actuator. The physical
constraint to consider is $x_1>0$, while the equilibrium points of
the system are
$$x_2^*=0,\quad x_3^*=(c_0+x_1^*)\sqrt{2c_1x_1^*(a_1+a_2{x_1^*}^2)},$$
The aim of controller design in this example is to stabilize the
system in $x_1^*>0$ equilibrium point. Hence, let us consider the
following desired interconnection matrix
$$J_d=\begin{bmatrix}
0 & 1 & 0 \\
-1 & 0 & \alpha(x) \\
0 & -\alpha(x) & 0
\end{bmatrix},$$
where $\alpha$ is a design parameter and $R_d=R$. The corresponding
PDEs are given as
$$K_2=0,\quad -K_1-bK_2+\alpha K_3=-\alpha \frac{x_3}{c_1(x_1+c_0)},$$
in which $K_i$s are considered as defined in (\ref{9.5}). The
corresponding Pfaffian differential equations are
$$\frac{dx_1}{-1}=\frac{dx_3}{\alpha}=\frac{dH_a}{-\alpha \frac{x_3}{c_1(x_1+c_0)}}$$
For simplicity and due to  physical constraint, consider
$\alpha=\frac{\beta(x_1+c_0)}{x_1}$. Hence, one should solve
\begin{align}
\frac{dx_1}{-x_1}=\frac{dx_3}{\beta(x_1+c_0)}=\frac{dH_a}{- \frac{\beta x_3}{c_1}}.
\label{20}
\end{align}
In the sequel, it is shown that
\begin{align*}
H_a&=\phi\Big(\beta x_1+\beta c_0\ln(x_1)+x_3,x_2\Big)\\&-\frac{1}{2c_0c_1}x_3^2-\frac{\beta}{c_0c_1}x_1x_3-\frac{\beta}{2c_0c_1}x_1^2-\frac{\beta}{c_1}x_1,
\end{align*}
is the solution of this Pfaffian differential equations.
In order to derive non-homogeneous solution, write:
\begin{align*}
\frac{dx_3}{\beta(x_1+c_0)}=\frac{dH_a}{- \frac{\beta x_3}{c_1}}=\frac{\frac{x_3}{c_0c_1}dx_3+dH_a}{\frac{\beta x_1x_3}{c_0c_1}}&\\=\frac{\frac{x_3}{c_0c_1}dx_3+dH_a+\frac{\beta x_3}{c_0c_1}dx_1}{0}.&
\end{align*}
Unfortunately, the last equation  does not satisfy condition
(\ref{cr}). To rectify this, let us add the term $\frac{\beta
    x_1}{c_0c_1}dx_3$ to it, which results in
\begin{align*}
\frac{dx_3}{\beta(x_1+c_0)}=\frac{dH_a}{- \frac{\beta x_3}{c_1}}=\frac{\frac{x_3+\beta x_1}{c_0c_1}dx_3+dH_a+\frac{\beta x_3}{c_0c_1}dx_1}{\frac{\beta x_1^2}{c_0c_1}+\frac{\beta x_1}{c_1}}
\end{align*}
Finally, one may reach to the following Pfaffian differential
equation
$$\frac{x_3+\beta x_1}{c_0c_1}dx_3+dH_a+\frac{\beta x_3}{c_0c_1}dx_1+\frac{\beta x_1}{c_0c_1}dx_1+\frac{\beta}{c_1}dx_1=0,$$
which has the following solution
$$H_a=-\frac{1}{2c_0c_1}x_3^2-\frac{\beta}{c_0c_1}x_1x_3-\frac{\beta}{2c_0c_1}x_1^2-\frac{\beta}{c_1}x_1.$$
The homogeneous solution of (\ref{20}) is derived easily as follows
$$H_a=\phi(\beta x_1+\beta c_0\ln(x_1)+x_3,x_2).$$
Thus, one can suitably define the constants such that $x^*$ becomes
a stable equilibrium point.

\subsection{ Third order food-chain system}
Consider the following model for third  order food-chain system
based on~\cite{nunna2015constructive} in PCH form (\ref{6}) with the
following values
\begin{align*}
J&=\begin{bmatrix}
0 & x_1x_2 & 0 \\ -x_1x_2 & 0 & x_2x_3 \\ 0 & -x_2x_3 & 0
\end{bmatrix},\hspace{2mm}D=\begin{bmatrix}
x_1 & 0 & 0\\ 0 & x_2 & 0 \\ 0 & 0 & x_3
\end{bmatrix}\\
g&=[0,0,1]^T, \qquad H=x_1+x_2+x_3
\end{align*}
where $x_i$ denotes the population  of $i$-th species.
In~\cite{ortega2000stabilization} it is shown that the PDE (\ref{3})
is not solvable with $J_d=J$ and $R_d>0$ since the span of the first
2 rows of $J_d-R_d$ is not involutive. This matching equation with
the  following matrices
$$J_d=\begin{bmatrix}
0 & J_1 & J_2 \\ -J_1 & 0 & J_3 \\ -J_2 & -J_3 & 0
\end{bmatrix},D=\begin{bmatrix}
R_1 & 0 & 0\\ 0 & R_2 & 0 \\ 0 & 0 & R_3
\end{bmatrix},H_d=H+H_a,$$
is in the following form
\begin{align*}
&\resizebox{.99\hsize}{!}{$-x_1+x_1x_2=-R_1(1+K_1)+J_1(1+K_2)+J_2(1+K_3),$}\\
&-x_2-x_1x_2+x_2+x_3=-J_1(1+K_1)-R_2(1+K_2)\\&\hspace{60mm}+J_3(1+K_3),
\end{align*}
in which $K_i$s are defined in (\ref{9.5}). If we set $R_d=I$ as the
simplest choice,  it is inferred that with
$J_1=0,J_2=f(x_1),J_3=g(x_2)$ the PDEs are involutive. This means
that homogeneous part of PDEs has a solution. The overall of PDEs
has also a solution if non-homogeneous solution of a PDE satisfies
the other. The corresponding Pfaffian differential equations are
\begin{align*}
&\frac{dx_1}{-1}=\frac{dx_2}{0}=\frac{dx_3}{f(x_1)}=\frac{dH_a}{-x_1+x_1x_2+1-f(x_1)},\\
&\frac{dx_1}{0}=\frac{dx_2}{-1}=\frac{dx_3}{g(x_3)}=\frac{dH_a}{-x_2-x_1x_2+x_2x_3+1}.
\end{align*}
The solution of these equations using the explained methods is
\begin{align*}
&H_a=\phi_1(x_2,\int f(x_1)dx_1+x_3)+\frac{1}{2}x_1^2-\frac{1}{2}x_1^2x_2-x_1\\&\hspace{54mm}+\int f(x_1)dx_1,\\&
H_a=\phi_2(x_1,x_2+\int \frac{1}{g(x_3)}dx_3)+\frac{1}{2}x_2^2+\frac{1}{2}x_1x_2^2-x_2\\&\hspace{27mm}+\int \big(\alpha(x_2,x_3)+\beta(x_2,x_3)g(x_3)\big),
\end{align*}
where $\alpha$ and $\beta$ should be defined such that the last term
is integrable. Now we should define $f(x_1)$ and $g(x_2)$ such that
the non-homogeneous solution of a PDE lies in the homogeneous
solution of other PDE. Hence, by defining $f(x_1)=-1$ and
$f(x_3)=0$, the solution of PDE is
$$H_a=\phi(x_1-x_3)+\frac{1}{2}x_2^2+\frac{1}{2}x_2^2(x_1-x_3)-x_2.$$

\begin{figure}[t]
    \center
    \includegraphics[scale=.45]{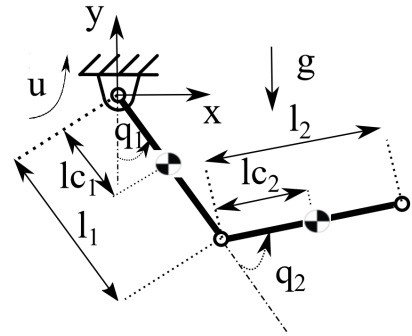}%
    \centering \caption{ Schematic of Pendubot. Merely the first joint is actuated. } \label{sh3}
\end{figure}
\subsection{Pendubot}
Here the IDA-PBC method is applied to pendubot system. The robot
consists of two revolute joints in which merely the first one is
actuated~\cite{wang2011unified}. The schematic of this system is
shown in Fig.~\ref{sh3}. The dynamic model of the robot may be
expressed in the form of (\ref{1a}) with the following
matrices~\cite{sandoval2008interconnection},
\begin{align}\label{16}
M&=\begin{bmatrix}
c_1+c_2+2c_3\cos(q_2) & c_2+c_3\cos(q_2) \\
c_2+c_3\cos(q_2) & c_2
\end{bmatrix}\\
G&=[1,0]^T,\quad V=-c_4g\cos(q_1)-c_5g\cos(q_1+q_2),
\nonumber
\end{align}
where the constants $c_i$s are given as follows
\begin{align*}
c_1&=m_1l_{c1}^2+m_2l_1^2+I_1,\quad c_2=m_2l_{c2}^2+I_2,\nonumber \\
c_3&=m_2l_1l_{c2},\quad c_4=m_1l_{c1}+m_2l_1,\quad c_5=m_2l_{c2}.
\end{align*}
In~\cite{sandoval2008interconnection}, it is shown that the
corresponding KE-PDE given in (\ref{4a}) is simplified to the
following equation for this system:
\begin{align}
&2c_3\sin(q_2)\big(\lambda_3^2+\lambda_3\lambda_4\big)
\nonumber\\&+\lambda_4\frac{d}{d
    q_2}\Big(\lambda_3\big(c_2+c_3\cos(q_2)\big) +\lambda_4c_2\Big)=0
\label{17}
\end{align}
in which $$M_dM^{-1}:=\begin{bmatrix}
\lambda_1 & \lambda_2 \\ \lambda_3 & \lambda_4
\end{bmatrix}.$$
Note that two other PDEs generated form KE-PDE (\ref{4a}) may be
solved by suitable definition of the matrix $J_2$. The PE-PDE
(\ref{5a}) for this system results in:
\begin{align}
\label{18}
\lambda_3\nabla_{q_1}V_d+\lambda_4\nabla_{q_2}V_d=c_5g\sin(q_1+q_2)
\end{align}
Since, PDE (\ref{17}) has two unknown variables, for simplicity, assume
that $\lambda_4=k\lambda_3$, and reduce it to the following
Pfaffian differential equations:
\begin{align*}
\resizebox{.99\hsize}{!}{$\frac{d q_1}{0}=\frac{d q_2}
{k\lambda_3(c_2+c_3\cos(q_2)+kc_2)}=\frac{d \lambda_3}
{-c_3\lambda_3^2\sin(q_2)(2+k)}$}
\end{align*}
Let us define $k=-1$ to simplify these equations. The
non-homogeneous solution is derived from the following equation
$$\frac{d \lambda_3}{\lambda_3}=\tan(q_2)d q_2,$$
which has the solution $\lambda_3=-\frac{1}{\cos(q_2)}$.  Note that
the homogeneous solution is trivially found to be $\phi(q_1)$. The
corresponding Pfaffian equations to PDE (\ref{18}) are given as
follows
\begin{align*}
\frac{d q_1}{-1}=\frac{d q_2}{1}=\frac{d V_d}{c_5g\cos(q_2)\sin(q_1+q_2)}
\end{align*}
The homogeneous solution is $V_d=\phi(q_1+q_2)$. In order to compute the
non-homogeneous solution, we should derive an equation in the form
of
\begin{align}
f_1(q_1,q_2)dq_1+f_2(q_1,q_2)dq_2+dV_d=0,
\label{19}
\end{align}
in which,
$$ -f_1+f_2=c_5g\cos(q_2)\sin(q_1+q_2),$$
and the following constraint resulted from (\ref{cr}) shall be
satisfied
$$\frac{\partial f_2}{\partial q_1}=\frac{\partial f_1}{\partial q_2}.$$
Combination of the two above equations yields to the following
equation
$$\frac{\partial f_2}{\partial q_1}-
\frac{\partial f_2}{\partial q_2}=-c_5g\cos(q_1+2q_2).$$ The
solution to this equation is $f_2=c_5g\sin(q_1+2q_2)$. Therefore,
the Pfaffian equation (\ref{19}) yields to
\begin{align*}
\hspace{1mm}\Big(c_5g\sin(q_1+2q_2)-c_5g\cos(q_2)\sin(q_1+q_2)\Big)dq_1&\\+c_5g\sin(q_1+2q_2)dq_2-dV_d=0.&
\end{align*}
Now one can apply the proposed procedure in section~\ref{s3} to
solve this equation. However, to make it short, rewrite it in the
following form
\begin{align*}
c_5g\sin(q_2)\cos(q_1+q_2)dq_1+\Big(c_5g\sin(q_2)\cos(q_1+q_2)&\\+c_5g\cos(q_2)\sin(q_1+q_2)\Big)dq_2-dV_d=0,&
\end{align*}
whose solution may be found easily as
$$V_d=c_5g\sin(q_1+q_2)\sin(q2).$$
\begin{remark}\normalfont
    In~\cite{sandoval2008interconnection}, the simplest solution to this
    problem is reported, in which $\lambda_3$ and $\lambda_4$
    are set to constant values. Here, a nontrivial solution is derived
    with enlarged domain of attraction.
    In~\cite{sandoval2008interconnection}, it is assumed that
    $q_2\in(-\epsilon,\epsilon)$ with
    $\epsilon=\arccos(\frac{c_2}{c_3})$. This limitation is also
    released in the proposed solution, where
    $q_2\in(-\frac{\pi}{2},\frac{\pi}{2})$.
\end{remark}
\begin{figure}[b]
    \center
    \includegraphics[scale=.4]{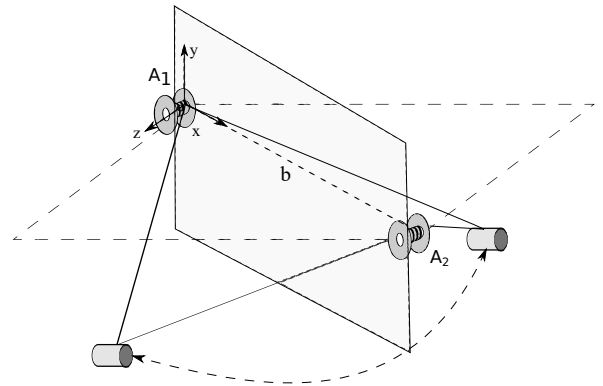}%
    \centering \caption{ Schematic of underactuated cable driven robots. The end-effector has a swing out of the vertical plane passes from actuators. } \label{sh1}
\end{figure}
\subsection{Underactuated spatial cable driven robot }
The schematic of this robot is shown in Fig.~\ref{sh1}. This system
is a planar robot which may have out-of-plane oscillation. Assume
that the center of coordinate is located on the first actuator, and
the position of actuators are given by:
\begin{equation}
\label{34}
A_1=\begin{bmatrix}
0 & 0 & 0
\end{bmatrix}^T\qquad \qquad
A_2=\begin{bmatrix}
b & 0 & 0
\end{bmatrix}^T
\end{equation}
Dynamic matrices of the robot may be easily found as
\begin{align}
M&=mI_3, \qquad\qquad\hspace{2mm} V=mgy \nonumber \\
G&=\begin{bmatrix}
\frac{x}{l_1} & \frac{x-b}{l_2} \\
\frac{y}{l_1} & \frac{y}{l_2} \\
\frac{z}{l_1} & \frac{z}{l_2}
\end{bmatrix}\qquad q=\begin{bmatrix}
x \\ y \\ z
\end{bmatrix}
\label{35}
\end{align}
where $q$ denotes the position of end-effector, $m$ denotes the
payload mass, and
\begin{align*}
l_1^2=x^2+y^2+z^2, \qquad l_2^2=(x-b)^2+y^2+z^2.
\end{align*}
Furthermore, assume that the cables are massless and infinitely
stiff. The equilibrium points of the robot
are $q^*=[x^*,y^*,0]$. Since these are natural equilibrium points of
the robot, one may use potential energy shaping for the controller
design. However, in this work we try to shape the total energy of
the system for a broader representation. For this robot, KE-PDE
introduced in (\ref{4a}) yields to:
\begin{align*}
G^\bot
\{-m^{-1}M_d\nabla_q(p^TM_d^{-1}p)+2J_2M_d^{-1}p\}=0,
\end{align*}
with
$$G^\bot=\begin{bmatrix}
0 & -bz & by
\end{bmatrix}.
$$
As explained in~\cite{acosta2005interconnection}, the general
solution of KE-PDE is obtained from the following equation
\begin{align}
\displaystyle\sum_{i=1}^{n} \gamma_i(q)\frac{d M_d^{-1}}{d q_i}=-[\mathcal{J}(q)A^T(q)+A(q)\mathcal{J}^T(q)]
\end{align}
where
\begin{align*}
J_2&=\begin{bmatrix}
0 & \tilde{p}^T\alpha_1 & \tilde{p}^T\alpha_2 & \dots & \tilde{p}^T\alpha_{n-1} \\ -\tilde{p}^T\alpha_1 & 0 & \tilde{p}^T\alpha_n & \dots & \tilde{p}^T\alpha_{2n-3} \\
\vdots&\vdots&\vdots&\ddots &\vdots\\-\tilde{p}^T\alpha_{n-1}& -\tilde{p}^T\alpha_{2n-3} &\dots & &0 \end{bmatrix},\\
\tilde{p}&=M^{-1}p,\quad
\mathcal{J}= [\alpha_1\vdots\alpha_2\vdots\cdots\vdots\alpha_{n_0}]\in\mathbb{R}^{n\times n_0},\\
A&=-[W_1(G^\bot)^T,\dots,W_{n_0}(G^\bot)^T]\in\mathbb{R}^{n_0\times n},\\
\gamma&=G^\bot M_dM^{-1}.
\end{align*}
In order to define $W_i$s, one may define matrices $F^{kl}$ with
$k,l\in\{1,...,n\}$ as follows
$$F^{kl}_{ij}=\begin{cases}
1 \quad \mbox{if}\quad j>i,i=k \hspace{1mm}\mbox{and}\hspace{1mm} j=l \\
0 \quad \mbox{otherwise}
\end{cases} $$
and set $W^{kl}=F^{kl}-(F^{kl})^T$ while $W_i$s as
$$W_1=W^{12},W_2=W^{13},...,W_{n_0}=W^{(n-1)n}.$$
By this means one should solve the following PDE:
\begin{align}
&(-zM_{d_{22}}+yM_{d_{23}})\frac{\partial M_d}{\partial y}+(-zM_{d_{23}}+yM_{d_{33}})\frac{\partial M_d}{\partial z}\nonumber\\&=m\begin{bmatrix}
2(-z\alpha_{1_1}+y\alpha_{2_1}) & *& *\\
-z\alpha_{2_1}+y(\alpha_{2_2}+\alpha_{3_1}) & 2y\alpha_{3_2} & *\\ y\alpha_{2_3}+z(\alpha_{3_1}-\alpha_{1_3}) & y\alpha_{3_3}+z\alpha_{3_2} & 2z\alpha_{3_3}
\end{bmatrix},
\end{align}
where the $*$'s in the last matrix denote that it is symmetric. It
is clear that $M_{d_{11}},M_{d_{12}},M_{d_{13}}$ may be found
arbitrary and the remaining terms shall satisfy the following
equations
\begin{align}
&(-zM_{d_{22}}+yM_{d_{23}})\frac{\partial M_{d_{22}}}{\partial y}+(-zM_{d_{23}}+yM_{d_{33}})\frac{\partial M_{d_{22}}}{\partial z}\nonumber\\&=2my\alpha_{3_2},\nonumber\\
&(-zM_{d_{22}}+yM_{d_{23}})\frac{\partial M_{d_{23}}}{\partial y}+(-zM_{d_{23}}+yM_{d_{33}})\frac{\partial M_{d_{23}}}{\partial z}\nonumber\\&=my\alpha_{3_3}+mz\alpha_{3_2},\nonumber \\ &(-zM_{d_{22}}+yM_{d_{23}})\frac{\partial M_{d_{33}}}{\partial y}+(-zM_{d_{23}}+yM_{d_{33}})\frac{\partial M_{d_{33}}}{\partial z}\nonumber\\&=2mz\alpha_{3_3}.\label{36}
\end{align}
This is a system of PDEs with two arbitrary function. Hence, it is
possible to convert it to a single PDE. However, there is no a simple
analytical solution for it.
%

Apply the proposed Theorem to find the solution for this PDE. In
order to convert  (\ref{36}) to Pfaffian equations, substitute first
and third equation of (\ref{36}) in the second equation. This yields
to:
\begin{align}
\label{38}
\frac{d y}{P_1}=\frac{d z}{P_2}=\frac{d M_{d_{23}}}{R}
\end{align}
with
\begin{align}
&P_1=-zM_{d_{22}}+yM_{d_{23}} \nonumber\\
&P_2=-zM_{d_{23}}+yM_{d_{33}} \nonumber\\
&\resizebox{.99\hsize}{!}{$R=-(\frac{y}{2}\frac{\partial M_{d_{33}}}{\partial z}+\frac{z^2}{2y}\frac{\partial M_{d_{22}}}{\partial z}-\frac{y^2}{2z}\frac{\partial M_{d_{33}}}{\partial y}-\frac{z}{2}\frac{\partial M_{d_{22}}}{\partial y})M_{d_{23}}$}\nonumber\\&+\frac{y}{2z}(-zM_{d_{22}}\frac{\partial M_{d_{33}}}{\partial y}+yM_{d_{33}}\frac{\partial M_{d_{33}}}{\partial z})\nonumber\\&+\frac{z}{2y}(-zM_{d_{22}}\frac{\partial M_{d_{22}}}{\partial y}+yM_{d_{33}}\frac{\partial M_{d_{22}}}{\partial z})\nonumber
\end{align}
where $M_{d_{22}}$ and $M_{d_{33}}$ are arbitrary functions.
Equation (\ref{38}) is equivalent to the following equation
\begin{align}
\label{39}
\frac{zd y+yd z}{-z^2M_{d_{22}}+y^2M_{d_{33}}}=\frac{d M_{d_{23}}}{R}
\end{align}
Note that the left hand side is independent of $M_{d_{23}}$, while
$R$ is summation of two terms including a linear term and an
independent term with respect to $M_{d_{23}}$. Pfaffian equation
(\ref{39}) is easier to solve if $R$ is independent of $M_{d_{23}}$.
Notice that the last two terms in $R$ are fractional and hard to be
used in the solution. In other words, notice that $M_d$ should be positive
definite. Hence, one may consider $M_{d_{22}}$ and $M_{d_{33}}$
as
$$M_{d_{22}}=\frac{y^2}{2}+k_1, \qquad\quad M_{d_{33}}=\frac{z^2}{2}+k_2$$
where $k_1,k_2>0$ to reduce the complexity. Substitute these values
in (\ref{39}):
\begin{align*}
\frac{zd y+yd z}{k_2y^2-k_1z^2}=\frac{d M_{d_{23}}}{k_2y^2-k_1z^2}
\end{align*}
The solution to this simplified equation is
$M_{d_{23}}=\frac{1}{2}yz$, hence, the structure of $M_d$ shall have
 the form of
\begin{align}
\label{40}
M_d=\begin{bmatrix}
* & * & * \\ * & \frac{y^2}{2}+k_1 & \frac{1}{2}yz \\
* & \frac{1}{2}yz & \frac{z^2}{2}+k_2
\end{bmatrix}
\end{align}
where undefined elements may be determined arbitrarily. Notice that
these elements do not appear in PE-PDE.

Potential energy PDE (\ref{4a}) for this robot may be derived as:
\begin{align*}
-bmgz&=bm^{-1}(-zM_{d_{22}}+yM_{d_{23}})\frac{\partial V_d}{\partial y}\nonumber\\&+bm^{-1}(-zM_{d_{23}}+yM_{d_{33}})\frac{\partial V_d}{\partial z}.
\end{align*}
Substitute (\ref{40}) in this equation to reach to:
\begin{align*}
-m^2gz=-k_1z\frac{\partial V_d}{\partial y}+k_2y\frac{\partial V_d}{\partial z}
\end{align*}
This is a simple PDE, that can be solved easily by Lemma~\ref{l1}.
The corresponding Pfaffian equations are
$$\frac{d x}{0}=\frac{d y}{-k_1z}=\frac{d z}{k_2y}=\frac{d V_d}{-m^2gz}$$
It is clear that $x=c_1$ and $k_2y^2+k_1z^2=c_2$ are the solutions
of the first two equalities. Thus, homogeneous solution of PDE is
given as
$$V_{d}=\Phi(x,k_2y^2+k_1z^2),$$
and from second and forth term, non-homogeneous solution is obtained as
$$V_{d}=\frac{m^2g}{k_1}(y-y^*).$$
\begin{remark}\normalfont
In this example, we have used total energy shaping method for the
spatial cable driven robot. Note that to the best of authors'
knowledge, none of the reported results on the topic of total energy
shaping of an underactuated robot,
e.g.~\cite{acosta2005interconnection,viola2007total,donaire2015shaping}
can be used to find the solution.
\end{remark}

\begin{figure}[b]
    \center
    \includegraphics[scale=.45]{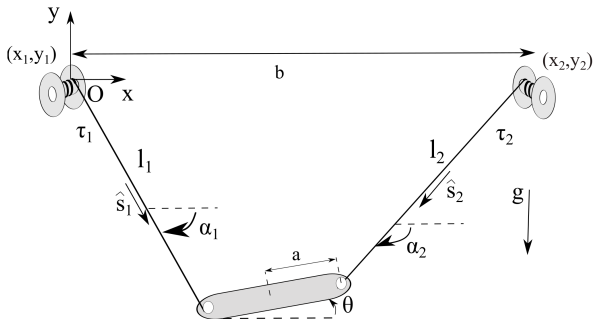}%
    \centering \caption{ Schematic of planar underactuated cable driven robots. The end-effector has a swing in the plane. } \label{sh2}
\end{figure}
\subsection{Underactuated planar cable driven robot}
In this Example let us apply IDA-PBC method to a 3-DOF underactuated
planar cable driven robot. The schematic of this robot is
illustrated in Fig.~\ref{sh2}. Dynamic matrices of the robot are in
the form of (\ref{1a})  as given in~\cite{harandi2019point}:
\begin{align*}
G^T=
\begin{bmatrix}
\frac{x-a\cos(\theta)}{l_1} & \frac{y-a\sin(\theta)}{l_1} &  \begin{array}{c} -a\cos(\theta)\frac{y-a\sin(\theta)}{l_1}\\ +a\sin(\theta)\frac{x-a\cos(\theta)}{l_1}\end{array} \\
\frac{x-b+a\cos(\theta)}{l_2} & \frac{y+a\sin(\theta)}{l_2} & \begin{array}{c}a\cos(\theta)\frac{y+a\sin(\theta)}{l_2}\\ -a\sin(\theta)\frac{x-b+a\cos(\theta)}{l_2}\end{array}
\end{bmatrix}
\end{align*}
\begin{align*}
M=\begin{bmatrix}
m & 0 & 0 \\ 0 & m & 0 \\ 0 & 0 & I
\end{bmatrix},
V=mgy,\hspace{1mm} q=\begin{bmatrix}
x \\ y \\ \theta
\end{bmatrix}
\end{align*}
For this robot, the manifold of equilibrium points may be derived
as:
\begin{align*}
G^\perp\nabla_q V=0 \quad \implies \quad
-2xy\cos(\theta)+by\cos(\theta)&\\+ab\sin(\theta)\cos(\theta)+2x^2\sin(\theta)-2bx\sin(\theta)=0.&
\end{align*}
As indicated in~\cite{harandi2019point}, these points are  natural
equilibrium points of the system; thus, only potential energy
shaping is required.

The PE-PDE (\ref{5a}) for this system is as follows:
\begin{align}
&\resizebox{.99\hsize}{!}{$\Big(-2xy\cos(\theta)+by\cos(\theta)+ab\sin(\theta)\cos(\theta)+2x^2\sin(\theta)$}\nonumber\\\nonumber&\resizebox{.99\hsize}{!}{$-2bx\sin(\theta)\Big)mga=a\Big(2\cos(\theta)y^2-2\sin(\theta)xy+by\sin(\theta)$}\\\nonumber&-ab\sin^2(\theta)\Big)\frac{\partial V_d}{\partial x}+a\Big(-2xy\cos(\theta)+by\cos(\theta)\\\nonumber&+ab\sin(\theta)\cos(\theta)+2x^2\sin(\theta)-2bx\sin(\theta)\Big)\frac{\partial V_d}{\partial y}\\&+\Big(2ax\sin(\theta)-2ay\cos(\theta)+by-ab\sin(\theta)\Big)\frac{\partial V_d}{\partial \theta}\label{42}
\end{align}
Finding the solution to this PDE is a prohibitive task. However,
it can be solved in a systematic way using the proposed method in
section~\ref{s3}. Corresponding Pfaffian differential equation is:
\begin{align}
\frac{d x}{P_1}=\frac{d y}{P_2}=\frac{d \theta}{P_3}=\frac{d V_d}{mgP_2}
\label{11}
\end{align}
with
\begin{align*}
&P_1=a\big(2\cos(\theta)y^2-2\sin(\theta)xy+by\sin(\theta)-ab\sin^2(\theta)\big)\\
&P_2=a\big(-2xy\cos(\theta)+by\cos(\theta)+ab\sin(\theta)\cos(\theta)\\&\hspace{5mm}+2x^2\sin(\theta)-2bx\sin(\theta)\big)\\
&P_3=\big(2ax\sin(\theta)-2ay\cos(\theta)+by-ab\sin(\theta)\big).
\end{align*}
To compute the homogeneous solution, let us derive a Pfaffian
equation that satisfies condition (\ref{cr}). For this purpose, and
considering (\ref{42}), it is reasonable to derive a Pfaffian
equation whose corresponding coefficients of $dx$ and $dy$ are
merely function of $\theta$. Hence, let's start with the following
expression to omit $x^2$ and $y^2$ from denominator
\begin{align}
\label{11.5}
\big(4a\cos(\theta)\big)dx+\big(4a\sin(\theta)\big)dy
+\big(-4a\sin(\theta)x&\nonumber\\+4a\cos(\theta)y\big)d\theta,&
\end{align}
which results in
\begin{align*}
&\resizebox{.99\hsize}{!}{$\frac{(4\cos(\theta))dx+(4\sin(\theta))dy+(-4a\sin(\theta)x+
        4a\cos(\theta)y)d\theta}{-4b\cos(\theta)y^2+4b\sin(\theta)xy+
        4sbx\sin^2(\theta)+4ab\sin(\theta)\cos(\theta)}$}\\&=Eq.(\ref{11})
\end{align*}
In this equation $x^2$ was omitted. To omit $y^2$, first add $-2bdx$
to (\ref{11.5}) and then add $2ab\sin(\theta)d\theta$ to it:
\begin{align*}
&\resizebox{.99\hsize}{!}{$\frac{(4a\cos(\theta))dx+(4a\sin(\theta))dy+(-4a\sin(\theta)x+
        4a\cos(\theta)y)d\theta}{0}$}\\&+\frac{-2bdx+2ab\sin(\theta)d\theta}{0}=Eq.(\ref{11})
\end{align*}
Thus, the nominator shall be zero, and by this means, one can easily
verify that condition (\ref{cr}) holds. Although solving the
Pfaffian equation
\begin{align}
\big(&4a\cos(\theta)-2b\big)dx+\big(4a\sin(\theta)\big)dy+\big(-4a\sin(\theta)x\nonumber\\&+4a\cos(\theta)y+2ab\sin(\theta)\big)d\theta=0,
\label{12}
\end{align}
is not hard, let us apply the procedure proposed in section~\ref{s3}
to find the solution in a systematic manner.
$$U\big((4a\cos(\theta)-2b)x+4a\sin(\theta)y\big)=C, \qquad \mu=1$$
$K$ is derived as
$$K=R+4a\sin(\theta)x-4a\cos(\theta)y=2ab\sin(\theta).$$
Finally, by using Lemma~\ref{l1}, the solution is given by
\begin{align*}
V_d=\phi\Big(U-2ab\cos(\theta)\Big)&=\Phi\Big(\big(4a\cos(\theta)-2b\big)x\\&+4a\sin(\theta)y-2ab\cos(\theta)\Big).
\end{align*}

With a similar approach, we try to get a separable Pfaffian equation in the following form
$$P(x_1)dx_1+Q(x_2)dx_2+R(x_3)dx_3=0,$$
which is easily integrable and has the following solution
$$\phi\bigg(\int P(x_1)dx_1+\int Q(x_2)dx_2+\int R(x_3)dx_3\bigg).$$
After some manipulations, the following equation is obtained
\begin{align}
\frac{xdx+ydy-\frac{b}{2}dx+\frac{ab}{2}\sin(\theta)d\theta}{0}=Eq.(\ref{11})
\label{13}
\end{align}
The solution of (\ref{13}) is
\begin{align*}
V_d=\phi\big(x^2+y^2-\frac{b}{2}x-\frac{ab}{2}\cos(\theta)\big).
\end{align*}
Notice that since we are shaping the potential energy in here, the
non-homogenious solution is equal to the open loop potential energy,
i.e. $V_d=mgy$.

\begin{remark}\normalfont
    The fist impression of PDE (\ref{42}) is very inconvenient, and
    finding its solution is a prohibitive task, to the best of author's
    knowledge not being reported in the literature and it is not possible to solve it using any software. The power of
    proposed method to restate and reformulate this problem to some
    Pfaffian differential equation is the key point to solve this
    challenging problem.
\end{remark}

\section{Conclusions}
In this paper, we derived suitable solution to the PDEs arising in
controller design methods such as in IDA-PBC. By using the Sneddon's
method, a first-order PDE is represented by some equivalent Pfaffian
differential equations. It was shown that if integrability condition
holds for a Pfaffian differential equation with three variables,
then the solution could be easily found. In order to illustrate how
this method can be applied in practice, it was implemented to a
number of different benchmark systems through which the IDA-PBC are
designed. Although, the systems being investigated in this paper
include magnetic levitation system, pendubot and two underactuated
cable driven robots, the application of the proposed method is
general and is not limited to these case studies.

\bibliography{ref}   
\bibliographystyle{IEEEtran}

\end{document}